\documentclass[12pt]{article}
\usepackage[dvips]{epsfig}

\begin{document}
\begin{flushright}
OHSTPY-HEP-T-98-011 \\
hep-th/9806133
\end{flushright}
\vspace{20mm}
\begin{center}
{\LARGE The DLCQ Spectrum of ${\cal N} =(8,8)$  Super Yang-Mills  }
\\
\vspace{20mm}
{\bf F.Antonuccio, O. Lunin, S.Pinsky} \\
\vspace{4mm}
{\em Department of Physics,\\ The Ohio State University,\\ Columbus,
OH 43210, USA\\
\vspace{8mm}
{\bf H.-C. Pauli, S.Tsujimaru} \\
\vspace{4mm}
Max-Planck-Institut f\"{u}r Kernphysik, \\ 69029 Heidelberg, Germany}

\end{center}
\vspace{20mm}
\begin{abstract}

We consider the $1+1$ dimensional ${\cal N} = (8,8)$
supersymmetric matrix field
theory obtained from a dimensional reduction of ten dimensional 
${\cal N} = 1$  super Yang-Mills.
The gauge groups we consider are  
U($N$) and SU($N$), where $N$ is finite but arbitrary. 
We adopt light-cone coordinates, and choose to work in the light-cone gauge.
Quantizing this theory via Discretized Light-Cone Quantization (DLCQ) 
introduces an integer, $K$, which restricts
the light-cone momentum-fraction of constituent quanta to be
 integer multiples of
$1/K$. Solutions to the DLCQ bound state equations are 
obtained for $K=2,3$ and $4$ by 
discretizing the light-cone super charges,
which preserves supersymmetry manifestly.
We discuss degeneracies in the massive spectrum that appear
to be independent of
the light-cone compactification,   
and are therefore expected to be present in the decompactified limit
$K \rightarrow \infty$.  
Our numerical results also support the claim that the 
SU($N$) theory has a mass gap.

\end{abstract}
\newpage

\baselineskip .25in

\section{Introduction}

The non-perturbative properties
of super Yang-Mills theories
have received a lot of attention lately.
In a seminal paper by Witten \cite{witt95}, it
was shown that the low energy dynamics of 
$N$ coincident D$p$-branes could be described by $p+1$
dimensional U($N$) super Yang-Mills.
This insight was instrumental in motivating 
the M(atrix) theory conjecture \cite{bfss97},
and also played a role in the AdS/CFT correspondence
recently proposed by Maldacena \cite{mald97}. 

In summary, theorists are now grappling with
the rather surprising fact that Yang-Mills theories
seem to know more about the dynamics of string theory than 
previously conceived. Moreover, physics
in many space-time dimensions may be described consistently by 
low dimensional Yang-Mills theories. There
is therefore renewed interest in studying the non-perturbative 
properties of low dimensional super Yang-Mills theories.

Motivated by these developments, we consider the 
$1+1$  dimensional supersymmetric matrix field
theory obtained from a dimensional reduction of ten dimensional 
${\cal N} = 1$  super Yang-Mills, which results in a two dimensional
gauge theory with ${\cal N} = (8,8)$ supersymmtery. The
possible gauge groups are U($N$) and SU($N$), where $N$ is finite but
arbitrary. A similar theory with
${\cal N} = (1,1)$ supersymmtery was studied recently 
in \cite{alp98}.

After introducing light-cone coordinates, and adopting
the light-cone gauge, it is a straightforward
procedure to implement Discrete Light-Cone Quantization (DLCQ) 
in order to extract numerical bound state 
solutions \cite{pb85}. As was pointed out in the earlier work \cite{sakai95},
exact supersymmtery may be preserved in the DLCQ spectrum 
if we choose to discretize the light-cone supercharges rather than the
light-cone Hamiltonian.  

The complexity of the ${\cal N} = (8,8)$ 
model far exceeds any other two dimensional 
theories studied in the context of DLCQ 
(see \cite{bpp98} for an extensive review), since there are now 
eight boson and eight fermion fields that propagate as physical 
modes. In practice, this means we
can only probe the theory for rather crude discretizations
($K\leq 4$, where $1/K$ is the smallest unit of
light-cone momentum). Despite this shortcoming,
we are able to 
resolve some interesting features of the decompactified
($K \rightarrow \infty$) theory. 
In particular, we are able to count degeneracies 
of certain states in the massive spectrum, and 
establish evidence for the existence
of a mass gap in the SU($N$) theory. 

The organization of the paper may be summarized as follows; in 
Section \ref{dlcqformulation} we
introduce the $1+1$ dimensional ${\cal N} = (8,8)$
supersymmetric gauge theory, which we formulate in light-cone coordinates.
Explicit expressions for the quantized light-cone supercharges are 
written down, followed by a discussion on the 
DLCQ formulation of the theory. 
In Section \ref{numericalresults} we
tabulate the results of our numerical DLCQ analysis,
highlighting the degeneracies observed in the spectrum.
We also argue why the numerical results are 
consistent with the existence of a mass gap; i.e. 
there are no {\em normalizable} massless states in 
the SU($N$) theory other than the light-cone vacuum. 
A summary of our observations, and further discussion,
appears in Section \ref{conclusions}.
The formulation
of ten-dimensional super Yang-Mills theory in light-cone 
coordinates is presented in Appendix \ref{ymills10}.

\section{Light-Cone Quantization and DLCQ at Finite $N$}
\label{dlcqformulation}
The two dimensional ${\cal N} = (8,8)$
supersymmetric gauge theory we are interested in
may be formally obtained 
by dimensionally reducing $9+1$ dimensional ${\cal N}=1$
super Yang-Mills to $1+1$ dimensions. For the sake of completeness,
we review the underlying ten dimensional light-cone Yang-Mills theory
in Appendix \ref{ymills10} -- in perhaps more detail
than is customary -- although the ideas should be 
familiar to many
readers.  

Dimensional reduction of the ten dimensional Yang-Mills
action (\ref{LCversion}) given in  Appendix \ref{ymills10}
is carried out by
stipulating that all fields are independent of the (eight) 
transverse
coordinates\footnote{The space-time points
in ten dimensional Minkowski space are parametrized by coordinates
$(x^0,x^1,\dots,x^9)$.} 
$x^I$, $I=1,\dots,8$. 
We may therefore assume
that the fields depend only on the light-cone variables $\sigma^{\pm}
= \frac{1}{\sqrt{2}}(x^0 \pm x^9)$. The resulting two dimensional
theory may be described by the action
\begin{eqnarray}
S_{1+1}^{LC} & = & \int d\sigma^+ d\sigma^- \hspace{1mm}
 \mbox{tr} \left( \frac{1}{2}D_\alpha X_I D^\alpha X_I + \frac{g^2}{4}
            [X_I,X_J]^2 - \frac{1}{4} F_{\alpha \beta} F^{\alpha \beta} 
 \right. \nonumber \\
& & \hspace{20mm}
+ \hspace{1mm}
{\rm i} \theta_R^T D_+ \theta_R +   {\rm i}\theta_L^T D_- \theta_L 
    - \sqrt{2}g\theta_L^T \gamma^I[X_I,\theta_R] \left. \frac{}{} \right),  
\label{LCversionreduced}
\end{eqnarray}
where the repeated indices $\alpha,\beta$ are summed over light-cone
indices $\pm$, and $I,J$ are summed over transverse
indices $1,\dots,8$.
The eight scalar fields $X_I(\sigma^+,\sigma^-)$ represent
$N \times N$ Hermitian matrix-valued fields, and are 
remnants of
the transverse components of the
ten dimensional gauge field $A_\mu$, while $A_{\pm}(\sigma^+,\sigma^-)$ 
are the
light-cone gauge field components of the residual 
two dimensional U($N$) or SU($N$) gauge symmetry. 
The spinors $\theta_R$ and
$\theta_L$ are remnants of
the right-moving and left-moving
projections of a sixteen component real spinor in the ten dimensional
theory. The components of  $\theta_R$ and $\theta_L$
transform in the adjoint representation of the gauge group.  
  $F_{\alpha \beta} =  
\partial_{\alpha} A_\beta - \partial_\beta A_\alpha
    +{\rm i}g[A_\alpha, A_\beta]$ is the two dimensional
gauge field curvature tensor, while
$D_\alpha =  \partial_\alpha + {\rm i}g[A_\alpha,\cdot]$ is the covariant
derivative for the (adjoint) spinor fields.
The eight $16 \times 16$ real symmetric matrices
$\gamma^I$ are defined in Appendix \ref{ymills10}.

Since we are working in the light-cone frame, it is natural
to adopt the light-cone gauge $A_- = 0$. With
this gauge choice, the action (\ref{LCversionreduced}) becomes   
\begin{eqnarray}
{\tilde S}_{1+1}^{LC}&=&
\int d\sigma^+d\sigma^- {\rm {tr}} \Bigg(\partial_+X_I\partial_-X_I +
{\rm i} 
\theta_R^T\partial_+ \theta_R + {\rm i}\theta_L^T\partial_- \theta_L 
\nonumber\\
&+&\frac{1}{2}(\partial_-A_+)^2 +gA_+J^+ 
-\sqrt{2}g \theta_L^T \gamma^I [X_I, \theta_R ] 
+\frac{g^2}{4}[X_I, X_J ]^2\Bigg), 
\label{EQ6}
\end{eqnarray}
where $J^+ ={\rm i}[X_I, \partial_-X_I]+2\theta_R^T\theta_R$ 
is the longitudinal
momentum current.
The (Euler-Lagrange) equations of motion for the $A_+$
and $\theta_L$ fields are now
 \begin{eqnarray}
&&\partial_-^2A_+=gJ^+, \label{firstc}\\
&& \sqrt2 {\rm i}\partial_-\theta_L=g\gamma^I [X_I,\theta_R].
\label{secondc}  
\end{eqnarray}
These are evidently constraint equations, since
they are independent of the light-cone time $\sigma^+$.
The ``zero mode'' of the constraints above provide
us with the conditions
\begin{equation}
\int d\sigma^- J^+=0, \mbox{     and      } 
\int d\sigma^- \gamma^I [X_I,\theta_R] =0,
\label{EQ4}
\end{equation}    
which will be imposed on the Fock space 
to select the physical states in the quantum theory. 
The first constraint above is well known in the literature,
and projects out the colorless states in
the quantized theory\cite{dak93}. The second (fermionic) constraint is
perhaps lesser well known, but certainly provides  non-trivial
relations governing the small-$x$ behavior of light-cone 
wave functions\footnote{If we introduce a mass term, such
relations become crucial in establishing finiteness 
conditions. See \cite{abd97}, for example.} \cite{abd97}. 
  
At any rate, equations (\ref{firstc}),(\ref{secondc}) permit one
to eliminate the non-dynamical fields $A_+$ and $\theta_L$  
in the theory, which is a particular feature of
light-cone gauge theories. There are no ghosts.
We may therefore write down explicit  expressions for the
light-cone momentum $P^+$ and Hamiltonian $P^-$ in terms
of the physical degrees of freedom of the theory, which
are denoted by the eight 
scalars $X_I$, and  right-moving spinor $\theta_R$: 
\begin{eqnarray}
P^+&=&\int d\sigma^- \hspace{1mm}
\mbox{tr}
\left( \partial_-X_I\partial_-X_I+{\rm i}
\theta_R^T \partial_-\theta_R \right), \label{P+}
\\  
P^- &=&g^2 
\int d\sigma^- {\rm {tr}}\Bigg(-\frac{1}{2} J^+\frac{1}{\partial_-^2}J^+
-\frac{1}{4}[X_I, X_J ]^2 \nonumber \\ 
&&\hspace{15mm}+\frac{{\rm i}}{2} 
(\gamma^I [X_I, \theta_R])^T
\frac{1}{\partial_-} \gamma^J [X_J, \theta_R]\Bigg). 
\label{P-}
\end{eqnarray}
The light-cone Hamiltonian propagates a given field configuration
in light-cone time $\sigma^+$, and contains all the non-trivial
dynamics of the interacting field theory. 

In the representation for the $\gamma^I$ matrices specified by 
(\ref{gamma9}) in Appendix \ref{ymills10}, we may write
\begin{equation}
\theta_R = { u \choose 0}, \label{spin8} 
\end{equation}
where $u$ is an eight component real spinor.      

In terms of their Fourier modes, the fields 
may be expanded at light-cone time $\sigma^+=0$ to give\footnote{
The symbol $\dagger$ denotes quantum conjugation, and does not
transpose matrix indices.} 
\begin{eqnarray}
&&X^I_{pq}(\sigma^-)= \frac{1}{\sqrt{2\pi}}
\int_{0}^{\infty}\frac{dk^+}{\sqrt{2 k^+}}\Big(a^I_{pq}(k^+)
e^{-{\rm i}k^+\sigma^-}
+ {a^I_{qp}}^{\dagger}(k^+)e^{{\rm i}k^+\sigma^-}\Big), 
\hspace{4mm} I=1,\dots,8; \hspace{3mm} \label{Xexp}\\
&&u^{\alpha}_{pq}(\sigma^-)=\frac{1}{\sqrt{2 \pi}}\int_0^{\infty}
\frac{dk^+}{\sqrt{2}} 
\Big(b^{\alpha}_{pq}(k^+)e^{-{\rm i}k^+\sigma^-}
+ {b^\alpha_{qp}}^{\dagger}(k^+)e^{{\rm i}k^+\sigma^-}\Big), 
\hspace{4mm} \alpha=1,\dots,8. \label{uexp}
\end{eqnarray}
For the gauge group U($N$), the (anti)commutation relations take the form
\begin{eqnarray}
&&[a^I_{pq}(k^+), {a^J_{rs}}^{\dagger}(k'^+)]=
\delta^{IJ}\delta_{pr}\delta_{qs} 
\delta(k^+- k'^+), \\
&&\{ b^{\alpha}_{pq}(k^+), {b^{\beta}_{rs}}^{\dagger}(k'^+)\}=
\delta^{\alpha\beta}
\delta_{pr}\delta_{qs}\delta(k^+- k'^+),
\end{eqnarray}
while for SU($N$), we have the corresponding relations 
\begin{eqnarray}
&&[a^I_{pq}(k^+), {a^J_{rs}}^{\dagger}(k'^+)]=
\delta^{IJ}(\delta_{pr}\delta_{qs} - \frac{1}{N}\delta_{pq} \delta_{rs}) 
\delta(k^+- k'^+), \\
&&\{ b^{\alpha}_{pq}(k^+), {b^{\beta}_{rs}}^{\dagger}(k'^+)\}=
\delta^{\alpha\beta}
(\delta_{pr}\delta_{qs}
- \frac{1}{N}\delta_{pq} \delta_{rs})
\delta(k^+- k'^+).
\end{eqnarray}
An important simplification of the light-cone quantization is that 
the light-cone vacuum  is the Fock vacuum $\vert 0 \rangle$, defined by   
\begin{equation}
a^I_{pq}(k^+)\vert 0 \rangle =b^{\alpha}_{pq}(k^+)\vert 0 \rangle=0, 
\end{equation}
for all positive longitudinal momenta $k^+ > 0$. 
We therefore have $P^+\vert 0 \rangle= P^-\vert 0 \rangle=0$.
  
The ``charge-neutrality'' condition (first integral constraint
from (\ref{EQ4})) requires that all the color indices must be 
contracted for physical states.
Thus physical states are formed by color traces of the 
boson and fermion creation operators 
${a^I}^{\dagger},{b^{\alpha}}^{\dagger}$
acting on the light-cone vacuum.  A single trace of these
creation operators may be identified
as a single closed string, where each creation operator
(or `parton'), carrying some longitudinal
momentum $k^+$,  represents a
 `bit' of the string. A product of traced operators
is then a multiple string state, and the quantity $1/N$
is analogous to a string coupling constant.

\medskip

At this point, we may determine explicit expressions
for the quantized light-cone operators $P^{\pm}$ by substituting
the mode expansions (\ref{Xexp}),(\ref{uexp}) into 
equations (\ref{P+}),(\ref{P-}). The mass operator
$M^2 \equiv 2 P^+ P^-$ may then be diagonalized to solve for the bound 
state mass spectrum.
However, as was pointed out in \cite{sakai95},
it is more convenient to determine the quantized expressions
for the supercharges, since this leads to a regularization prescription
for $P^-$ that preserves supersymmetry even in the discretized theory.

In order to elaborate upon this last remark, first note that
the continuum theory possesses sixteen supercharges,
which may be derived from the dimensionally reduced 
form of the ten dimensional ${\cal N} = 1$ supercurrent:
\begin{eqnarray}
 Q^+_{\alpha} & = & 2^{1/4} \int_{-\infty}^{\infty} d\sigma^- \hspace{1mm}
\mbox{tr} \left( \partial_- X_I \cdot \beta_{I \eta \alpha} \cdot u_{\eta} 
     \right) \label{Q+}\\
 Q^-_{\alpha} & = & g \int_{-\infty}^{\infty} d\sigma^- \hspace{1mm}
\mbox{tr} \left( -2^{3/4} \cdot J^+  \frac{1}{\partial_-} 
              u_{\alpha} +  2^{-1/4} {\rm i} [X_I,X_J] \cdot 
(\beta_I \beta_J^T)_{\alpha \beta} \cdot u_{\beta}  \right),
\label{Q-}
\end{eqnarray}
where $\alpha=1,\dots,8$, and repeated indices are 
summed. The eight $8 \times 8$ real
matrices $\beta_I$ are discussed in Appendix \ref{ymills10}. By explicit
calculation or otherwise, these charges satisfy the following
relations:
\begin{eqnarray}
       \{ Q^+_{\alpha}, Q^+_{\beta} \} & = & 
   \delta_{\alpha \beta} \cdot \frac{1}{\sqrt{2}} P^+ 
\label{superQplus} \\
 \{ Q^-_{\alpha}, Q^-_{\beta} \} & = & 
   \delta_{\alpha \beta} \cdot \frac{1}{\sqrt{2}} P^-  \label{superQminus}
\end{eqnarray}
If we substitute the mode expansions (\ref{Xexp}),(\ref{uexp}) 
into equations (\ref{Q+}),(\ref{Q-}) 
for the light-cone supercharges $Q^{\pm}_{\alpha}$,
we obtain the following `momentum representations' for these charges:
\begin{eqnarray}
Q^+_{\alpha} & = & 2^{-3/4} {\rm i} \int_0^{\infty}
 dk \hspace{1mm} \sqrt{k} \cdot \beta_{I\eta \alpha} \cdot 
\left( a^{\dagger}_{Iij}(k) b_{\eta ij}(k) -
       b^{\dagger}_{\eta ij}(k) a_{I ij}(k) \right),
\label{Qplus}
\end{eqnarray}
and 
\begin{eqnarray}
\lefteqn{ Q^-_{\alpha} = \frac{{\rm i} 2^{-1/4} g}{\sqrt{\pi}}
  \int_0^{\infty} dk_1 dk_2 dk_3 \hspace{1mm} 
\delta( k_1 + k_2 - k_3) \cdot \left\{ \frac{}{} \right. }   & &  
\nonumber \\
& & 
\frac{1}{2\sqrt{k_1 k_2}} \left( \frac{k_2 - k_1}{k_3} \right)
\left[ b^{\dagger}_{\alpha i j}(k_3)a_{I i m}(k_1)a_{I m j}(k_2) -
      a_{I i m}^{\dagger}(k_1)a_{I m j}^{\dagger}(k_2)
b_{\alpha i j}(k_3) \right] \nonumber \\
& + & 
\frac{1}{2\sqrt{k_1 k_3}} \left( \frac{k_1 + k_3}{k_2} \right)
\left[ a^{\dagger}_{I i m}(k_1)b_{\alpha m j}^{\dagger}(k_2)a_{I i j}(k_3) -
      a_{I i j}^{\dagger}(k_3)a_{I i m}(k_1)
b_{\alpha m j}(k_2) \right] \nonumber \\
& + & 
\frac{1}{2\sqrt{k_2 k_3}} \left( \frac{k_2 + k_3}{k_1} \right)
\left[ a^{\dagger}_{I i j}(k_3)b_{\alpha i m}(k_1)a_{I m j}(k_2) -
      b_{\alpha i m}^{\dagger}(k_1)a_{I m j}^{\dagger}(k_2)
a_{I i j}(k_3) \right] \nonumber \\
& - & \frac{1}{k_1} \left[ 
b^{\dagger}_{\beta i j}(k_3)b_{\alpha i m}(k_1)b_{\beta m j}(k_2) +
      b_{\alpha i m}^{\dagger}(k_1) b_{\beta m j}^{\dagger}(k_2)
b_{\beta i j}(k_3) \right] \nonumber \\
& - & 
\frac{1}{k_2} \left[ 
b^{\dagger}_{\beta i j}(k_3)b_{\beta i m}(k_1)b_{\alpha m j}(k_2) +
      b_{\beta i m}^{\dagger}(k_1) b_{\alpha m j}^{\dagger}(k_2)
b_{\beta i j}(k_3) \right] \nonumber \\
& + &   
\frac{1}{k_3} \left[ 
b^{\dagger}_{\alpha i j}(k_3)b_{\beta i m}(k_1)b_{\beta m j}(k_2) +
      b_{\beta i m}^{\dagger}(k_1) b_{\beta m j}^{\dagger}(k_2)
b_{\alpha i j}(k_3) \right] \nonumber \\
& + & \hspace{8mm} (\beta_I \beta_J^T - \beta_J \beta_I^T )_{\alpha \beta}
\times \left( \frac{}{} \right. \nonumber \\
& & \frac{1}{4\sqrt{k_1 k_2}}
\left[ b^{\dagger}_{\beta i j}(k_3)a_{I i m}(k_1)a_{J m j}(k_2) +
      a_{J i m}^{\dagger}(k_1)a_{I m j}^{\dagger}(k_2)
b_{\beta i j}(k_3) \right] \nonumber \\
& + & 
\frac{1}{4\sqrt{k_2 k_3}}
\left[ a^{\dagger}_{J i j}(k_3)b_{\beta i m}(k_1)a_{I m j}(k_2) +
      b_{\beta i m}^{\dagger}(k_1)a_{J m j}^{\dagger}(k_2)
a_{I i j}(k_3) \right] \nonumber \\
& + &
\frac{1}{4\sqrt{k_3 k_1}}
\left[ a^{\dagger}_{I i j}(k_3)a_{J i m}(k_1)b_{\beta m j}(k_2) +
      a_{I i m}^{\dagger}(k_1)b_{\beta m j}^{\dagger}(k_2)
a_{J i j}(k_3) \right] \left. \frac{}{} \right) \left. \frac{}{}
\right\}, \label{Qminus}
\end{eqnarray}
where repeated indices are always summed: $\alpha,\beta = 1,\dots,8$
(SO(8) spinor indices), $I,J=1,\dots , 8$ (SO(8) vector indices),
and $i,j,m=1,\dots , N$ (matrix indices).

In order to implement the DLCQ formulation\footnote{
It might be useful to consult \cite{sakai95,dak93,anp97,pin97}
for an elaboration of DLCQ in models
with adjoint fermions.}
of the bound state problem -- which is tantamount
to imposing periodic boundary conditions 
$\sigma^- \sim \sigma^- + 2 \pi R$ --
we simply restrict the momentum variable(s) 
appearing in the expressions for $Q^{\pm}_{\alpha}$ 
(equations (\ref{Qplus}),(\ref{Qminus})) to the following
discretized set of
momenta: $\{ \frac{1}{K}P^+, \frac{2}{K}P^+,  \frac{3}{K}P^+, \dots \}$.
Here, $P^+$ denotes the total light-cone momentum of a state,
and may be thought of as a fixed constant,
since it is easy to form a Fock basis that is already diagonal 
with respect
to the quantum operator $P^+$ \cite{pb85}. The integer $K$ is called
the `harmonic resolution', and $1/K$ measures the coarseness of our 
discretization --
we recover the continuum by taking the limit $K \rightarrow \infty$. 
Physically, $1/K$ represents the smallest 
positive\footnote{We exclude the zero mode $k^+=0$ in our analysis;
the massive spectrum is not expected to be affected by this omission,
but there are issues concerning the light-cone vacuum that 
involve $k^+=0$ modes \cite{pin97a,mrp97}.}
unit of longitudinal momentum-fraction allowed for 
each parton in a Fock state.  

Of course, as soon as we implement the DLCQ procedure,
which is specified unambiguously by the harmonic resolution $K$,
the integrals appearing in the definitions for $Q^{\pm}_{\alpha}$
are replaced by finite sums, and the eigen-equation
$2 P^+ P^- |\Psi \rangle = M^2 |\Psi \rangle$ is reduced
to a finite matrix diagonalization problem. In this last
step, we use the fact that $P^-$ is proportional to the square
of any one of the eight supercharges $Q^-_{\alpha}$, $\alpha=1,\dots,8$
(equation (\ref{superQminus})),
and so the problem of diagonalizing $P^-$ is 
equivalent to diagonalizing any one of the 
eight supercharges $Q^-_{\alpha}$. As was pointed out
in \cite{sakai95}, this procedure yields a supersymmetric
spectrum for any resolution $K$.
In the present work, we are able to perform numerical diagonalizations
for $K=2,3$ and $4$ with the help of Mathematica and a desktop PC.
 
The fact that we may choose any one of the eight supercharges
to calculate the spectrum provides a strong test for
the correctness of our computer program. 
As expected, we find that the spectrum we obtain
by squaring the eigenvalues of any two different
supercharges yields the {\em same} massive spectrum.
Moreover, the spectrum turns out to be {\em exactly supersymmetric},
which is also what we require. Such tests are very convenient
when studying complicated 
models; for example, in the expression
for $Q^-_{\alpha}$ (eqn (\ref{Qminus})), there are approximately
3500 terms.

\section{DLCQ Bound State Solutions}
\label{numericalresults}
We consider discretizing the light-cone supercharge
$Q^-_{\alpha}$ for a particular $\alpha \in \{1,2,\dots,8\}$, and
for the values $K=2,3,4$.
 For a given resolution $K$,
the light-cone momenta of partons 
in a given Fock state must be some positive integer 
multiple of $P^+/K$, where $P^+$ is the total light-cone momentum of
the state.
 For example, when $K=2$, there are precisely 256 Fock states
in the U($N$) theory that are made up from two partons:
\begin{eqnarray}
\mbox{128 Bosons:} & &  
\left\{
\begin{array}{ll}
  \mbox{tr}[a^{\dagger}_I(\frac{1}{2}P^+)a^{\dagger}_J(\frac{1}{2}P^+)]
       |0\rangle &  I,J=1,2,\dots,8; \\
   \mbox{tr}[b^{\dagger}_{\alpha}(\frac{1}{2}P^+)
b^{\dagger}_{\beta}(\frac{1}{2}P^+)]
       |0\rangle & \alpha,\beta=1,2,\dots,8; 
\hspace{3mm} (\alpha \neq \beta);  \\
 \mbox{tr}[a^{\dagger}_I(\frac{1}{2}P^+)]
 \mbox{tr}[a^{\dagger}_J(\frac{1}{2}P^+)]
       |0\rangle &  I,J=1,2,\dots,8; \\
   \mbox{tr}[b^{\dagger}_{\alpha}(\frac{1}{2}P^+)]
   \mbox{tr}[b^{\dagger}_{\beta}(\frac{1}{2}P^+)]
       |0\rangle & \alpha,\beta=1,2,\dots,8;
                   \hspace{3mm} (\alpha \neq \beta);  
       \end{array}
      \right. \\
& & \nonumber \\ 
\mbox{128 Fermions:} & &  
\left\{
\begin{array}{ll}
  \mbox{tr}[a^{\dagger}_I(\frac{1}{2}P^+)b^{\dagger}_{\alpha}
    (\frac{1}{2}P^+)]
       |0\rangle &  I,\alpha=1,2,\dots,8; \\
   \mbox{tr}[a^{\dagger}_{I}(\frac{1}{2}P^+)]
\mbox{tr}[b^{\dagger}_{\alpha}(\frac{1}{2}P^+)]
       |0\rangle & I,\alpha=1,2,\dots,8;   \\
\end{array}
      \right.
\end{eqnarray}
Of course, there are an additional 16 single particle
states: eight bosons of the form 
$\mbox{tr}[a^{\dagger}_I(P^+)]|0\rangle$ and eight fermions
of the form 
$\mbox{tr}[b^{\dagger}_{\alpha}(P^+)]|0\rangle$.
This gives a total of $128+8$ bosons and $128+8$ fermions
in the DLCQ Hilbert space for the U($N$) theory.
If we calculate the matrix
representation of $Q^-_{\alpha}$ (for any $\alpha$) with
respect to this finite basis, we find that the masses
$M^2 \sim (Q^-_{\alpha})^2$ of all these states
are zero. In fact, this is what we expect.
First of all, it can be shown that the 
the light-cone supercharge $Q^-_{\alpha}$ for the U($N$)
gauge group is identical to the expression for the SU($N$)
supercharge. This is tantamount to saying that 
the U(1) part of the U($N$) theory 
decouples completely as a free field theory, and 
is identically zero for the light-cone Hamiltonian.
The U(1) states in the U($N$) DLCQ Fock space are readily
identified; they are precisely those states that
are made from a product of one-particle Fock states.
The remaining states -- consisting
of 64 bosons and 64 fermions -- belong to the SU($N$) Fock space,
and must therefore be single-trace states of two partons.
Since the supercharge changes the number of partons 
in a Fock state by one, it must annihilate 
any SU($N$) Fock state, which can only have 
two partons when $K=2$.

The decoupling of the U(1) degrees of freedom 
in the U($N$) theory provides trivial examples
of massless states,  
and implies that all the non-trivial
dynamics is contained in the SU($N$) gauge
theory. In particular,
investigating
the existence (or not) of massless states in the SU($N$) 
theory is a highly non-trivial problem for $K \geq 3$.
We will therefore restrict our attention to the SU($N$)
gauge theory.

To begin, we list all two parton states in the 
SU($N$) gauge theory for $K=3$:
\begin{eqnarray}
\mbox{128 Bosons:} & &  
\left\{
\begin{array}{ll}
  \mbox{tr}[a^{\dagger}_I(\frac{1}{3}P^+)a^{\dagger}_J(\frac{2}{3}P^+)]
       |0\rangle &  I,J=1,2,\dots,8; \\
   \mbox{tr}[b^{\dagger}_{\alpha}(\frac{1}{3}P^+)
b^{\dagger}_{\beta}(\frac{2}{3}P^+)]
       |0\rangle & \alpha,\beta=1,2,\dots,8;   
       \end{array}
      \right. \\
& & \nonumber \\ 
\mbox{128 Fermions:} & &  
\left\{
\begin{array}{ll}
  \mbox{tr}[a^{\dagger}_I(\frac{1}{3}P^+)b^{\dagger}_{\alpha}
    (\frac{2}{3}P^+)]
       |0\rangle &  I,\alpha=1,2,\dots,8; \\
 \mbox{tr}[a^{\dagger}_I(\frac{2}{3}P^+)b^{\dagger}_{\alpha}
    (\frac{1}{3}P^+)]
       |0\rangle &  I,\alpha=1,2,\dots,8. 
      \end{array}      
\right.
\end{eqnarray}
Thus, there are 128 bosons and 128 fermions that consist
of two partons. For three parton states,
where the momentum is shared equally among each parton, 
the states take the following form:
\begin{eqnarray}
\mbox{688 Bosons:} & &  
\left\{
\begin{array}{ll}
  \mbox{tr}[a^{\dagger}_I(\frac{1}{3}P^+)a^{\dagger}_J(\frac{1}{3}P^+)
      a^{\dagger}_K(\frac{1}{3}P^+)]
       |0\rangle &  I,J,K=1,2,\dots,8; \\
   \mbox{tr}[a^{\dagger}_{I}(\frac{1}{3}P^+)
b^{\dagger}_{\alpha}(\frac{1}{3}P^+)
b^{\dagger}_{\beta}(\frac{1}{3}P^+)
]|0\rangle & I,\alpha,\beta=1,2,\dots,8;   
       \end{array}
      \right. \\
& & \nonumber \\ 
\mbox{688 Fermions:} & &  
\left\{
\begin{array}{ll}
   \mbox{tr}[b^{\dagger}_{\alpha}(\frac{1}{3}P^+)
b^{\dagger}_{\beta}(\frac{1}{3}P^+)
      b^{\dagger}_{\gamma}(\frac{1}{3}P^+)]
       |0\rangle &  \alpha,\beta,\gamma=1,2,\dots,8; \\
   \mbox{tr}[a^{\dagger}_{I}(\frac{1}{3}P^+)
a^{\dagger}_{J}(\frac{1}{3}P^+)
b^{\dagger}_{\alpha}(\frac{1}{3}P^+)
]|0\rangle & I,J,\alpha=1,2,\dots,8.   
      \end{array}      
\right.
\end{eqnarray}
More specifically, there are 176 boson states of the 
form\footnote{
We use Polya Theory to count these states; 
we think of a necklace with three beads, where each bead may
be colored in eight distinct ways. The permutation
symmetry involving only rotations is  ${\bf Z}_3$, and the
`cyclic
index polynomial' is therefore $\frac{1}{3}[x_1^3 + 2x_3]$.
Hence there are $\frac{1}{3}[8^3 + 2\cdot 8]=176$ distinct
configurations modulo cyclic rotations.}
$\mbox{tr}[a^{\dagger}_I(\frac{1}{3}P^+)a^{\dagger}_J(\frac{1}{3}P^+)
      a^{\dagger}_K(\frac{1}{3}P^+)]
       |0\rangle$, and $8\times 8 \times 8 = 512$ states
of the form 
$\mbox{tr}[a^{\dagger}_{I}(\frac{1}{3}P^+)
a^{\dagger}_{J}(\frac{1}{3}P^+)
b^{\dagger}_{\alpha}(\frac{1}{3}P^+)
]|0\rangle$. Therefore, the 
SU($N$) $K=3$ DLCQ Hilbert space    
consists of 816 bosons and 816 fermions. It is indeed satisfying to find
that our computer algorithm generates precisely this number of states.  
The results of our DLCQ numerical diagonalization of $(Q^-_{\alpha})^2$
is summarized in Table \ref{K3masses}.  To test our numerical
algorithms, we diagonalize different supercharges, and find
the same spectrum -- which is consistent with supersymmetry.
\begin{table}[h!]
\begin{center}
\begin{tabular}{|c|c|}
\hline
\multicolumn{2}{|c|}{Bound State Masses $M^2$ for $K=3$ } \\
\hline
$M^2$ & Mass Degeneracy  \\
\hline
0 & $560+560$  \\
\hline
18 & $128+128$ \\
\hline 
72 & $112+112$  \\
\hline
126 & $16+16$  \\
\hline 
\end{tabular}
\caption{SU($N$) bound state masses $M^2$ 
in units $g^2 N/\pi$ for resolution $K=3$.
When expressed in these units, the masses are
independent of $N$ (i.e. there are no $1/N$
corrections at this resolution), and so these results are applicable
for any $N > 1$.
The notation `$128+128$'  above implies 
a 256--fold mass degeneracy in the spectrum with
128 bosons and 128 fermions. 
  \label{K3masses}}
\end{center}
\end{table}

Let us now consider resolution $K=4$. 
For the sake of definiteness, we enumerate 
carefully the SU($N$) DLCQ Fock space.
Firstly, bosonic Fock states with only two partons 
take the following form:
\begin{eqnarray}
\mbox{192 bosons (2 partons):} & &  
\left\{
\begin{array}{ll}
  \mbox{tr}[a^{\dagger}_I(\frac{1}{4}P^+)a^{\dagger}_J(\frac{3}{4}P^+)]
       |0\rangle &  I,J=1,2,\dots,8; \\
   \mbox{tr}[b^{\dagger}_{\alpha}(\frac{1}{4}P^+)
b^{\dagger}_{\beta}(\frac{3}{4}P^+)]
       |0\rangle & \alpha,\beta=1,2,\dots,8; \\
  \mbox{tr}[a^{\dagger}_I(\frac{2}{4}P^+)a^{\dagger}_J(\frac{2}{4}P^+)]
       |0\rangle &  I,J=1,2,\dots,8; \\  
 \mbox{tr}[b^{\dagger}_{\alpha}(\frac{2}{4}P^+)
b^{\dagger}_{\beta}(\frac{2}{4}P^+)]
       |0\rangle & \alpha,\beta=1,2,\dots,8;  \hspace{3mm} 
(\alpha \neq \beta); 
       \end{array} \nonumber
   \right.
\end{eqnarray}
It is straightforward to verify that there are 
$64+64+36+28=192$ such states.
Similarly, bosonic Fock states with three partons 
take the form
\begin{eqnarray}
\mbox{2048 bosons (3 partons):} & &  
\left\{
\begin{array}{ll}
  \mbox{tr}[a^{\dagger}_I(\frac{1}{4}P^+)a^{\dagger}_J(\frac{1}{4}P^+)
      a^{\dagger}_K(\frac{2}{4}P^+)]
       |0\rangle &  I,J,K=1,2,\dots,8; \\
   \mbox{tr}[a^{\dagger}_{I}(\frac{1}{4}P^+)
b^{\dagger}_{\alpha}(\frac{1}{4}P^+)
b^{\dagger}_{\beta}(\frac{2}{4}P^+)
]|0\rangle & I,\alpha,\beta=1,2,\dots,8; \\
  \mbox{tr}[a^{\dagger}_{I}(\frac{1}{4}P^+)
b^{\dagger}_{\alpha}(\frac{2}{4}P^+)
b^{\dagger}_{\beta}(\frac{1}{4}P^+)
]|0\rangle & I,\alpha,\beta=1,2,\dots,8; \\
 \mbox{tr}[a^{\dagger}_{I}(\frac{2}{4}P^+)
b^{\dagger}_{\alpha}(\frac{1}{4}P^+)
b^{\dagger}_{\beta}(\frac{1}{4}P^+)
]|0\rangle & I,\alpha,\beta=1,2,\dots,8,    
       \end{array}
      \right. \nonumber
\end{eqnarray}
and it is easily shown that there are $4\times 8^3 = 2048$ such states.
Enumerating all four parton bosonic Fock states
requires additional effort. Firstly, we consider
all single-trace bosonic Fock states with four partons;
these are listed below:
\begin{small}
\begin{eqnarray}
\mbox{8192 bosons (4 partons):} & &  
\left\{
\begin{array}{ll}
  \mbox{tr}[a^{\dagger}_I(\frac{1}{4}P^+)a^{\dagger}_J(\frac{1}{4}P^+)
      a^{\dagger}_K(\frac{1}{4}P^+)
     a^{\dagger}_L(\frac{1}{4}P^+)]
       |0\rangle &  I,J,K,L=1,\dots,8; \\
   \mbox{tr}[a^{\dagger}_{I}(\frac{1}{4}P^+)
a^{\dagger}_{J}(\frac{1}{4}P^+)
b^{\dagger}_{\alpha}(\frac{1}{4}P^+)
b^{\dagger}_{\beta}(\frac{1}{4}P^+)
]|0\rangle & I,J,\alpha,\beta=1,\dots,8; \\
  \mbox{tr}[a^{\dagger}_{I}(\frac{1}{4}P^+)
b^{\dagger}_{\alpha}(\frac{1}{4}P^+)
a^{\dagger}_{J}(\frac{1}{4}P^+)
b^{\dagger}_{\beta}(\frac{1}{4}P^+)
]|0\rangle & I,J,\alpha,\beta=1,\dots,8; \\
 \mbox{tr}[b^{\dagger}_{\alpha}(\frac{1}{4}P^+)
b^{\dagger}_{\beta}(\frac{1}{4}P^+)
b^{\dagger}_{\gamma}(\frac{1}{4}P^+)
b^{\dagger}_{\delta}(\frac{1}{4}P^+)
]|0\rangle & \alpha,\beta,\gamma,\delta=1,\dots,8.    
       \end{array}
      \right. \nonumber
\end{eqnarray}
\end{small}
The total number of such states is 8192, and decomposes 
as follows; there are 1044 states of the first type
listed above\footnote{We use Polya theory as before:
The cyclic permutation symmetry of a necklace with four beads,
each of which can be colored in eight distinct ways, is ${\bf Z}_4$,
and gives rise to the cyclic index polynomial 
$\frac{1}{4}[x_1^4+x_2^2+2x_4]$. Thus, there
are $\frac{1}{4}[8^4+8^2+2\cdot 8]=1044$ distinct configurations},
$8\times 8\times 8\times 8 = 4096$ 
states of the second type, 2016 states of the third
type\footnote{The symmetry here is the subgroup ${\bf Z}_2$ of ${\bf Z}_4$,
and the resulting cyclic index polynomial is $\frac{1}{2}[
x_1^4+x_2^2]$. Thus, there are $\frac{1}{2}[8^4+8^2]=
2080$ distinct states modulo cyclic permutations.
However, 64 of these states have zero norm, and
may be identified as those states
for which $(I,\alpha)=(J,\beta)$. Subtracting these states,
we are left with $2080-64=2016$ distinct states of the third type.},
and finally, 1036 states of the fourth 
type\footnote{The counting here is the same as in the first type
because of the ${\bf Z}_4$ cyclic symmetry, but we must also
subtract zero-norm states, which are precisely those 
states with $\alpha=\beta=\gamma=\delta$. There can only
be 8 such states, and so we have $1044-8=1036$ distinct
states overall.}.

The remaining four-parton bosonic Fock states are formed from
a product of two two-parton Fock states:
\begin{small}
\begin{eqnarray}
\mbox{4096 bosons (4 partons):} & &  
\left\{
\begin{array}{ll}
  \mbox{tr}[a^{\dagger}_I(\frac{1}{4}P^+)a^{\dagger}_J(\frac{1}{4}P^+)]
   \mbox{tr}[a^{\dagger}_K(\frac{1}{4}P^+)
     a^{\dagger}_L(\frac{1}{4}P^+)]
       |0\rangle &  I,J,K,L=1,\dots,8; \\
   \mbox{tr}[a^{\dagger}_{I}(\frac{1}{4}P^+)
a^{\dagger}_{J}(\frac{1}{4}P^+)]
\mbox{tr}[b^{\dagger}_{\alpha}(\frac{1}{4}P^+)
b^{\dagger}_{\beta}(\frac{1}{4}P^+)
]|0\rangle & I,J,\alpha,\beta=1,\dots,8; \\
  \mbox{tr}[a^{\dagger}_{I}(\frac{1}{4}P^+)
b^{\dagger}_{\alpha}(\frac{1}{4}P^+)]
\mbox{tr}[a^{\dagger}_{J}(\frac{1}{4}P^+)
b^{\dagger}_{\beta}(\frac{1}{4}P^+)
]|0\rangle & I,J,\alpha,\beta=1,\dots,8; \\
 \mbox{tr}[b^{\dagger}_{\alpha}(\frac{1}{4}P^+)
b^{\dagger}_{\beta}(\frac{1}{4}P^+)]
\mbox{tr}[b^{\dagger}_{\gamma}(\frac{1}{4}P^+)
b^{\dagger}_{\delta}(\frac{1}{4}P^+)
]|0\rangle & \alpha,\beta,\gamma,\delta=1,\dots,8.    
       \end{array}
      \right. \nonumber
\end{eqnarray}
\end{small}
Straightforward counting techniques yield 
666 states of the first type listed above, 1008 states of the second 
type, 2016 states of the third type, and 406 states of the fourth type,
giving a total of 4096 bosons. 

We therefore conclude that there are 10432 single-trace 
bosonic Fock states, and 4096 double-trace bosonic Fock states,
yielding 14528 bosons in total.

\medskip

We now enumerate all the fermions, which turns out to 
be a much simpler calculation. To begin, all two-parton
fermionic states have the form
\begin{eqnarray}
\mbox{192 fermions (2 partons):} & &  
\left\{
\begin{array}{ll}
  \mbox{tr}[a^{\dagger}_I(\frac{1}{4}P^+)b^{\dagger}_{\alpha}
(\frac{3}{4}P^+)]
       |0\rangle &  I,\alpha=1,2,\dots,8; \\
  \mbox{tr}[a^{\dagger}_I(\frac{3}{4}P^+)b^{\dagger}_{\alpha}
(\frac{1}{4}P^+)]
       |0\rangle &  I,\alpha=1,2,\dots,8; \\
  \mbox{tr}[a^{\dagger}_I(\frac{2}{4}P^+)b^{\dagger}_{\alpha}
(\frac{2}{4}P^+)]
       |0\rangle &  I,\alpha=1,2,\dots,8,
       \end{array} \nonumber
   \right.
\end{eqnarray}
and it is straightforward to check that there are $64+64+64=192$
such states. Note that this equals the number of two-parton bosonic
states. The enumeration of all three-parton fermionic
states is listed below:
\begin{eqnarray}
\mbox{2048 fermions (3 partons):} & &  
\left\{
\begin{array}{ll}
  \mbox{tr}[a^{\dagger}_I(\frac{1}{4}P^+)a^{\dagger}_J(\frac{1}{4}P^+)
      b^{\dagger}_{\alpha}(\frac{2}{4}P^+)]
       |0\rangle &  I,J,\alpha=1,\dots,8; \\
   \mbox{tr}[b^{\dagger}_{\alpha}(\frac{1}{4}P^+)
b^{\dagger}_{\beta}(\frac{1}{4}P^+)
b^{\dagger}_{\gamma}(\frac{2}{4}P^+)
]|0\rangle & \alpha,\beta, \gamma =1,\dots,8; \\
  \mbox{tr}[a^{\dagger}_{I}(\frac{2}{4}P^+)
a^{\dagger}_{J}(\frac{1}{4}P^+)
b^{\dagger}_{\alpha}(\frac{1}{4}P^+)
]|0\rangle & I,J,\alpha =1,\dots,8; \\
 \mbox{tr}[a^{\dagger}_{I}(\frac{1}{4}P^+)
a^{\dagger}_{J}(\frac{2}{4}P^+)
b^{\dagger}_{\alpha}(\frac{1}{4}P^+)
]|0\rangle & I,J,\alpha = 1,\dots,8, 
       \end{array}
      \right. \nonumber
\end{eqnarray}
and it is easy to verify that there are $4\times 8^3=2048$ such states.
Once again, this precisely matches the number of three-parton bosonic
states. Four-parton fermionic states may consist of
a single trace or a product of two traces. The single-trace Fock
states take the form
\begin{small}
\begin{eqnarray}
\mbox{8192 fermions (4 partons):} & &  
\left\{
\begin{array}{ll}
  \mbox{tr}[a^{\dagger}_I(\frac{1}{4}P^+)a^{\dagger}_J(\frac{1}{4}P^+)
      a^{\dagger}_K(\frac{1}{4}P^+)
     b^{\dagger}_{\alpha}(\frac{1}{4}P^+)]
       |0\rangle &  I,J,K,\alpha=1,\dots,8; \\
 \mbox{tr}[a^{\dagger}_I(\frac{1}{4}P^+)b^{\dagger}_{\alpha}
(\frac{1}{4}P^+)
      b^{\dagger}_{\beta}(\frac{1}{4}P^+)
     b^{\dagger}_{\gamma}(\frac{1}{4}P^+)]
       |0\rangle &  I,\alpha,\beta, \gamma =1,\dots,8,   
       \end{array}
      \right. \nonumber
\end{eqnarray}
\end{small}
and there are $2 \times 8^4 = 8192$ such states. 
This number agrees exactly with the number of single-trace
bosonic states with four partons, although we recall that
the counting of bosonic states was significantly more complicated.

Finally, four-parton fermionic states
with two traces take the form
\begin{small}
\begin{eqnarray}
\mbox{4096 fermions (4 partons):} & &  
\left\{
\begin{array}{ll}
  \mbox{tr}[a^{\dagger}_I(\frac{1}{4}P^+)a^{\dagger}_J(\frac{1}{4}P^+)]
   \mbox{tr}[a^{\dagger}_{K}(\frac{1}{4}P^+)
     b^{\dagger}_{\alpha}(\frac{1}{4}P^+)]
       |0\rangle &  I,J,K,\alpha=1,\dots,8; \\
   \mbox{tr}[a^{\dagger}_{I}(\frac{1}{4}P^+)
b^{\dagger}_{\alpha}(\frac{1}{4}P^+)]
\mbox{tr}[b^{\dagger}_{\beta}(\frac{1}{4}P^+)
b^{\dagger}_{\gamma}(\frac{1}{4}P^+)
]|0\rangle & I,\alpha,\beta,\gamma=1,\dots,8.   
       \end{array}
      \right. \nonumber
\end{eqnarray}
\end{small}
One may now verify that there are $36\times 64=2304$ states
of the first type, and $64 \times 28=1792$ states of
the second type, yielding 4096 states overall. 
This of course agrees with the number of double-trace bosonic
states calculated earlier.

We have thus verified that there are precisely an equal number
of bosons and fermions in the $K=4$ DLCQ Hilbert space of
the SU($N$) theory. The total number of states is
precisely $14528+14528 = 29056$. This reflects an important
feature of DLCQ; namely, {\em DLCQ preserves supersymmetry}.

We remark here that the computer
algorithm we use for constructing the DLCQ Fock states
involves choosing an arbitrary set of input Fock states, and then
repeatedly acting on this set by a preassigned number of supercharges 
until no new states are formed. These supercharges may then
be diagonalized on this sub-space of Fock states.
It is reassuring to find that this algorithm generates precisely
the number of states that we counted above. 

In order to determine the bound state spectrum,
we need to diagonalize a particular supercharge $Q^-_{\alpha}$
on the DLCQ Hilbert space. Fortunately, because of the sixteen
supersymmetries, 
we can reduce the problem of diagonalizing a 
$29056 \times 29056$ matrix to the problem of
diagonalizing sixteen $1816 \times 1816$ block matrices.
These block matrices may be reduced further; the double-trace
states are already diagonal with
respect to the mass-squared operator $M^2$,
and are massless, so they decouple from
the dynamics of single-trace
Fock states. Therefore, the block matrix involving only
single trace Fock states has dimensions  $1304 \times 1304$,
and is easily handled by a desk top PC.
  
The results of our numerical diagonalizations are presented
in Table \ref{K4masses}. Note that there are
$4096+4096$ massive states; for $K=3$, there were 
$256+256$ massive bound states.
\begin{table}[h!]
\begin{center}
\begin{tabular}{|c|c|}
\hline
\multicolumn{2}{|c|}{Bound State Masses $M^2$ for $K=4$ } \\
\hline
$M^2$ & Mass Degeneracy  \\
\hline
0 & $10432+10432$  \\
\hline
24 & $560+560$ \\
\hline 
29.668 & $128+128$  \\
\hline
32 & $432+432$  \\
\hline
$53.0605^{\ast}$ & $128+128$ \\ 
\hline 
56 & $16+16$ \\
\hline
72 & $768+768$ \\
\hline 
73.7982 & $16+16$ \\
\hline 
80 & $768+768$ \\
\hline
88 & $336+336$ \\
\hline
$90.3875^{\ast}$ & $112 + 112$ \\
\hline 
96 & $336+336$ \\
\hline
114.332 & $128+128$ \\
\hline 
120 & $112+112$ \\
\hline 
141.612 & $112 + 112$ \\
\hline
$151.091^{\ast}$ & $16+16$ \\
\hline
157.606 & $128+128$ \\
\hline
\end{tabular}
\caption{SU($N$) bound state masses $M^2$ 
in units $g^2 N/\pi$ for resolution $K=4$.
When expressed in these units, the masses are
independent of $N$ (i.e. there are no $1/N$
corrections at this resolution), and so these results are applicable
for any $N > 1$. Masses labeled with ${}^{\ast}$ correspond
to the states observed at the lower resolution $K=3$
(Table \ref{K3masses}). To
make this identification, it is necessary to study the 
Fock state expansion of these bound states.
  \label{K4masses}}
\end{center}
\end{table}
  
\section{Discussion}
\label{conclusions}
It is evident from the DLCQ bound state masses
summarized in Tables \ref{K3masses} and \ref{K4masses}
that there are a large number of massless states.
At first, this seems to be at odds with the claim
that the SU($N$) gauge theory is expected to have 
a mass gap \cite{witt95}. However, to determine whether
there is a mass gap or not, we need to investigate 
whether there are normalizable states with zero mass  
in the {\em continuum 
limit} $K \rightarrow \infty$. In our present study, we
only considered the values $K=2,3$ and $4$, and so it would 
seem hopeless at first to make any statements about the 
continuum theory. It turns out, however, that
there is already suggestive evidence of a mass gap
which can be obtained at these low resolutions.

The crucial observation is that all the massless states
in the DLCQ spectrum are made up of partons carrying
the smallest positive unit of light-cone momentum allowed
at the given resolution. For
example, at $K=2$, we saw that the SU($N$) 
Hilbert space consisted of two-parton Fock
states -- 64 bosons and 64 fermions (all massless) -- where 
each parton 
carried the smallest integer unit of light-cone momentum.
For $K=3$, we find that all the massless states
are a superposition of only three-parton
Fock states, so each parton carries
one unit of light-cone momentum. The states made from a superposition
of two-parton Fock states, which were massless at $K=2$, acquire a 
mass at the higher resolution $K=3$. 
Similarly, after studying carefully the DLCQ bound
states at resolution $K=4$, we find that 
the massless states are superpositions of only
four-parton Fock states.
Each parton in these Fock states carries precisely
one unit of light-cone momentum. 
There are no massless states
involving Fock states with two or three partons at $K=4$, so the 
massless states we observe at $K=2$ and $K=3$ have evidently acquired
a mass at the higher resolution. 

This pattern is very suggestive; namely,
we expect that at a given resolution $K$, the massless
states in the DLCQ spectrum will be a superposition
of {\em only} $K$-parton Fock states, so that each parton carries a
single unit of light-cone momentum. It is clear, then,
that as we take the continuum limit $K \rightarrow \infty$,
these massless states do not converge to any well-defined
massless state in the continuum, which contrasts
what is observed in a two dimensional supersymmetric model with
$(1,1)$ supersymmetry \cite{alp98}. Of course, this 
assumption
is not enough 
to establish the existence of a mass gap, since it is possible
that lighter massive states may appear at higher resolutions,
and possibly converge to zero in the limit $K \rightarrow \infty$ 
\cite{alp98}.
However, we note that the lightest massive states
at $K=4$ are heavier than the ones observed at $K=3$, and so 
increasing the resolution
does not appear to introduce lighter massive states.
Evidently, it would be desirable to probe larger
values of $K$ to help clarify this issue, and we leave this
for future work. Nevertheless, our results clearly support
the existence of a mass gap 
in the continuum SU($N$) supersymmetric gauge theory.

There is also additional information 
about the continuum theory that emerges from 
our DLCQ results. First of all,
the massive states observed at $K=3$
(see Table \ref{K3masses}) are also
observed at $K=4$ (Table \ref{K4masses})
with the same mass degeneracy. 
We therefore expect these degeneracies to be preserved 
for all values of $K$, including the  
continuum limit $K \rightarrow \infty$.
Our numerical results therefore indicate mass degeneracies 
that are expected to be present in the spectrum
of the continuum theory. 

We finally comment on possible connections between the  
DLCQ ${\cal N}=(8,8)$ model studied here and various
string-related
models. It
has already been claimed that at resolution $K$ 
one finds massless states made up of $K$-parton
Fock states,
so that each parton carries precisely one unit
of light-cone momentum. 
 If one thinks of $K$ as being large 
but finite, then these states become string-like states
made up of many `bits'.
One also finds that the lightest massive states 
at $K=4$ are composed of mainly three and four-parton
Fock states, and so, in general, one expects
the low energy spectrum to be dominated by string-like states --
a property that is in fact observed for two dimensional $(1,1)$
super Yang-Mills \cite{alp98}. This suggests that the DLCQ model
studied here might be closely related to the `string-bit' models originally
proposed by Thorn \cite{thorn}. Perhaps more intriguing
is the possible connection 
with matrix string theory \cite{dvv}.
In the DLCQ model we compactify a light-like direction, while for
matrix string theory, one works with the same Lagrangian, but
chooses instead to compactify a space-like coordinate, which
originates from the geometry of closed strings in Type IIA string theory.
It would be very interesting to compare these two schemes,
and possibly relate them. Perhaps understanding the
origin of quantized electric flux in the context of light-cone
quantized gauge theories \cite{pin97a,mrp97} 
will pave the way to a better understanding
of the significance of the DLCQ model studied here and
the dynamics of non-perturbative string theory.

\medskip
\begin{large}
 {\bf Acknowledgments}
\end{large}

F.A. is grateful to Jungil Lee for assistance with computer work.
S.T. is grateful for hospitality during his visit at Ohio State. 
\appendix
\section{Appendix: Super Yang-Mills in Ten Dimensions}
\label{ymills10}
Let's start with ${\cal N}=1$ super Yang-Mills theory 
in 9+1 dimensions with gauge group U($N$):
\begin{equation}
S_{9+1}=\int d^{10}x \hspace{1mm} \mbox{tr} \Bigg
(-\frac{1}{4} F_{\mu\nu}F^{ \mu\nu}+\frac{{\rm i}}{2} 
\bar{\Psi}\Gamma^{\mu}D_{\mu}\Psi\Bigg) , 
\label{EQ1}
\end{equation} 
where 
\begin{eqnarray}
F_{\mu\nu}&=&\partial_{\mu}A_{\nu}-\partial_{\nu}A_{\mu}
+{\rm i}g[A_{\mu},  A_{\nu}] , \\
D_{\mu}\Psi &=& \partial_{\mu}\Psi+{\rm i}g[A_\mu, \Psi],  
\end{eqnarray} 
and $\mu,\nu = 0,\dots,9$.
The Majorana spinor  $\Psi$ 
transforms in the adjoint representation of  U($N$). 
The (flat) space-time metric
$g_{\mu \nu}$ has signature $(+,-,\dots,-)$, and we adopt 
the normalization $\mbox{tr}(T^aT^b) = \delta^{a b}$ for
the generators of the U($N$) gauge group.

In order to realize the ten dimensional Dirac algebra  
$\{\Gamma_\mu, \Gamma_\nu\}=2g_{\mu\nu}$ in terms
of Majorana matrices, 
we use as building blocks the 
reducible ${\bf 8}_s + {\bf 8}_c$ representation
of the spin(8) Clifford Algebra. In block form, we have 
\begin{equation}
\gamma^I=\left(\begin{array}{cc}
0 & \beta_I\\
\beta_I^T & 0 
\end{array}\right), \hspace{7mm} I=1,\dots,8,
\end{equation} 
where the $8 \times 8$ real matrices, $\beta_I$, satisfy
$\{\beta_I,\beta_J^T \} = 2\delta_{IJ}$. This automatically
ensures the spin(8) algebra $\{\gamma^I,\gamma^J \} = 2\delta^{IJ}$
for the $16 \times 16$ real-symmetric matrices $\gamma^I$.
An explicit representation for the $\beta_I$ algebra 
may be given in terms of a tensor product of Pauli matrices \cite{schwarz}. 
In the present context, we may choose a representation such that
a ninth matrix, $\gamma^9 = \gamma^1 \gamma^2 \cdots \gamma^8$, which
anti-commutes with the other eight $\gamma^I$'s, 
takes the explicit form   
\begin{equation} 
\gamma^9=\left(\begin{array}{cc}
{\bf 1}_{8} & 0\\
0 & -{\bf 1}_{8} \end{array}\right). \label{gamma9}
\end{equation}
We may now construct $32 \times 32$ pure imaginary
(or Majorana) matrices $\Gamma^\mu$ which realize
the Dirac algebra for the Lorentz group SO($9,1$): 
\begin{eqnarray}   
&& \Gamma^0=\sigma_2 \otimes {\bf 1}_{16}, \\
&& \Gamma^I={\rm i}\sigma_1 \otimes \gamma^I, \hspace{6mm} I=1,\dots,8;\\
&& \Gamma^9= {\rm i}\sigma_1 \otimes \gamma^9.
\end{eqnarray}
The Majorana spinor therefore has 32 real components, and
since it transforms in the adjoint 
representation of U($N$),
each of these components may be viewed as an $N \times N$ 
Hermitian matrix. 

An additional matrix $\Gamma_{11}= 
\Gamma^0 \cdots \Gamma^9$, which is
equal to $\sigma_3\otimes {\bf 1}_{16}$ in the representation
specified by (\ref{gamma9}), 
is easily seen to anti-commute with all other gamma matrices, and
satisfies  $(\Gamma_{11})^2 = 1$. It is also real, and so 
the  Majorana spinor field $\Psi$ admits a chiral decomposition
via the projection operators $\Lambda_{\pm} \equiv 
\frac{1}{2}(1 \pm \Gamma_{11})$:
\begin{equation}
 \Psi = \Psi_+ + \Psi_-, \hspace{5mm}  \Psi_{\pm} =  \Lambda_{\pm} \Psi.
\end{equation}
We will therefore consider only spinors 
with positive chirality $\Gamma_{11} \Psi = +\Psi$ (Majorana-Weyl):
\begin{equation}
\Psi= 2^{1/4} { \psi \choose 0}, \label{spin16} 
\end{equation}         
where $\psi$ is a sixteen component real spinor, and the 
numerical factor $2^{1/4}$ is introduced for later convenience. 

Since $\gamma^9$ anti-commutes with the other eight $\gamma^I$'s,
and satisfies $(\gamma^9)^2 = 1$, we may construct further
projection operators $P_R \equiv \frac{1}{2}(1+\gamma^9)$ and
$P_L \equiv \frac{1}{2}(1-\gamma^9)$ which project out, respectively, the 
right-moving and left-moving components of the sixteen component
spinor $\psi$ defined in (\ref{spin16}):
\begin{equation}
 \psi = \psi_R + \psi_L, \hspace{5mm}  \psi_R =  P_R \psi, \hspace{3mm}
\psi_L =  P_L \psi.
\end{equation}
This decomposition is particularly useful when working
with light-cone coordinates, since in the light-cone gauge
one can express the left-moving component $\psi_L$ in terms of
the right-moving component $\psi_R$ by virtue of the
fermion constraint equation. We will derive this result shortly.
In terms of the usual ten dimensional Minkowski
space-time coordinates, the light-cone coordinates are given by
\begin{eqnarray}
 x^+ & = & \frac{1}{\sqrt{2}}(x^0 + x^9), \hspace{10mm}
\mbox{``time coordinate''} \\
 x^- & = & \frac{1}{\sqrt{2}}(x^0 - x^9), \hspace{10mm}
\mbox{``longitudinal space coordinate''}  \\ 
 {\bf x}^{\perp} & = & (x^1,\dots,x^8).
\hspace{13mm} \mbox{``transverse coordinates''} 
\end{eqnarray}
Note that the `raising' and `lowering' of the $\pm$ indices
is given by the rule $x^{\pm} = x_{\mp}$, 
while $x^I = -x_I$ for $I=1,\dots,8$,
as usual. It is now a routine task to demonstrate that
the Yang-Mills action (\ref{EQ1}) for the positive
chirality spinor (\ref{spin16}) is equivalent to
\begin{eqnarray}
S_{9+1}^{LC} & = & \int dx^+ dx^- d{\bf x}^{\perp} \hspace{1mm}
 \mbox{tr} \left( \frac{1}{2}F_{+-}^2 + F_{+I}F_{-I} - \frac{1}{4}F_{IJ}^2
 \right. \nonumber \\
& & \hspace{20mm}
+ \hspace{1mm}
{\rm i} \psi_R^T D_+ \psi_R +   {\rm i}\psi_L^T D_- \psi_L +
     {\rm i}\sqrt{2}\psi_L^T \gamma^I D_I \psi_R \left. 
\frac{}{} \right),
\label{LCversion}
\end{eqnarray}
where the repeated indices $I,J$ are summed over $(1,\dots,8)$.
Some surprising simplifications follow if we now choose 
to work in the {\em light-cone gauge} $A^+ = A_- = 0$. In
this gauge $D_- \equiv \partial_-$, and so the (Euler-Lagrange)
equation of motion for the left-moving field $\psi_L$
is simply 
\begin{equation}
 \partial_- \psi_L = -\frac{1}{\sqrt{2}}\gamma^I D_I \psi_R, 
\label{fermioncon}
\end{equation}
which is evidently a non-dynamical constraint equation, since it
is independent of the light-cone time. We may therefore eliminate
any dependence on $\psi_L$ (representing unphysical 
degrees of freedom) in favor of $\psi_R$, which carries the
eight physical fermionic degrees of freedom in the theory.    
In addition, 
the equation of motion for the $A_+$ field yields
Gauss' law:
\begin{equation} 
\partial_{-}^2 A_{+}=\partial_{-}\partial_{I}A_{I}+gJ^+
\label{apluscon}
\end{equation} 
where $J^+={\rm i}[A_{I},\partial_{-}A_{I}]+2\psi_{R}^T\psi_{R}$, and
so the $A_+$ field may also be eliminated
to leave the eight bosonic degrees of freedom $A_I$, $I=1,\dots,8$.
Note that the eight fermionic degrees of freedom
exactly match the eight
bosonic degrees of freedom associated with the transverse
polarization of a ten dimensional gauge field, which is of
course consistent
with the supersymmetry. We should emphasize that
unlike the usual 
covariant formulation of Yang-Mills, the light-cone formulation
here permits one to remove {\em explicitly} 
any unphysical degrees of freedom in the 
Lagrangian (or Hamiltonian); there 
are no ghosts. 


\vfil


\begin{thebibliography}{9999}
\bibitem{witt95} E.Witten, 
{\em Bound States Of Strings And $p$-Branes},
{\em Nucl.Phys.} {\bf B460}, (1996), 335-350,
 hep-th/9510135.
%
\bibitem{bfss97}T.Banks, W.Fischler, S.Shenker and L.Susskind,
{\em M Theory As A Matrix Model: A Conjecture},
 {\em Phys.Rev.} {\bf D55}, (1997), 5112-5128, hep-th/9610043.
%
\bibitem{mald97} Juan M. Maldacena,
{\em The Large N Limit of Superconformal Field Theories and Supergravity},
 hep-th/9711200.
%
\bibitem{alp98} F.Antonuccio, O.Lunin, S.Pinsky,
{\em Non-Perturbative Spectrum of Two Dimensional (1,1) Super 
Yang-Mills at Finite and Large $N$}, hep-th/9803170 
(to appear in {\em Phys.Rev.} {\bf D});
 F.Antonuccio, O.Lunin, S.Pinsky, {\em
Bound States of Dimensionally Reduced $\mbox{SYM}_{2+1}$ at Finite $N$},
hep-th/9803027, (To appear in {\em Phys.Lett.} {\bf B}).
%
\bibitem{pb85} H.-C. Pauli and S.J.Brodsky,
{\em Phys.Rev.} {\bf D32} (1985) 1993, 2001.
%
\bibitem{sakai95} Yoichiro Matsumura, Norisuke Sakai, Tadakatsu Sakai,
{\em  Mass Spectra of Supersymmetric Yang-Mills Theories in 
$1 + 1$ Dimensions}, hep-th/9504150,  {\em Phys. Rev.} 
{\bf D52} (1995) 2446. 
%
\bibitem{bpp98} S.J. Brodsky, H.C. Pauli, and S.S. Pinsky,
{\em Quantum Chromodynamics and Other Field Theories on the Light Cone}
(To appear in Phys.Rept.), hep-ph/9705477.
%
\bibitem{abd97} F.Antonuccio, S.Brodsky and S.Dalley,
{\em Light-cone Wavefunctions at Small $x$}, 
{\em Phys.Lett.} {\bf B412} (1997) 104-110. hep-ph/9705413.
%
\bibitem{dak93} S.Dalley and I.Klebanov,
 {\em String Spectrum of 1+1-Dimensional Large $N$ QCD with Adjoint Matter},
hep-th/9209049,  {\em Phys. Rev.} {\bf D47} (1993) 2517-2527;
G. Bhanot, K.Demeterfi and I. Klebanov,
{\em $1+1$-Dimensional Large $N$ QCD coupled to Adjoint Fermions},
 hep-th/9307111,  {\em Phys.Rev.} {\bf D48} (1993) 4980-4990. 
%
\bibitem{schwarz} M.B.Green, J.H.Schwarz, and E.Witten, {\em
 Superstring Theory}, Vol.1, CUP (1987).
%
\bibitem{anp97} F. Antonuccio, S.S. Pinsky,
{\em Phys.Lett} {\bf B397}:42-50,1997, hep-th/9612021.
%
\bibitem{pin97} S. Pinsky,
{\it ``The Analog of the t'Hooft Pion with Adjoint Fermions''}
Invited talk at New Nonperturbative Methods and Quantization of the Light
Cone, Les Houches.
France, 24 Feb - 7 Mar 1997. hep-th/9705242
%
\bibitem{pin97a} S. Pinsky,
{\em Phys.Rev} {\bf D56}:5040-5049,1997 hep-th/9612073
%
\bibitem{mrp97} S. Pinsky and D. Robertson,
{\em Phys.Lett } {\bf B 379} (1996) 169-178;
G. McCartor, D. G. Robertson and S. Pinsky
 {\em Phys.Rev} {\bf D56}:1035-1049,1997 hep-th/9612083
%
\bibitem{thorn}
C.B.Thorn, {\em Phys.Rev} {\bf D19} (1979) 639; C.B Thorn,
{\em Reformulating String Theory with the $1/N$ Expansion},
hep-th/9405069. 
%
\bibitem{dvv}L.Motl, {\em Proposals on Non-Perturbative 
Superstring Interactions}, hep-th/9701025;
T.Banks and N.Seiberg, {\em Strings from Matrices},
{\em Nucl.Phys.} {\em B497} (1997) 41, hep-th/9702187;
R.Dijkgraaf, E.Verlinde, H.Verlinde, {\em Matrix String Theory},
{\em Nucl.Phys.} {\bf B500} (1997) 43, hep-th/9703030.
\end{thebibliography}
\end{document}